\documentclass[
      letterpaper,
      prd,
      floatfix,
      preprint,
      superscriptaddress,
      nofootinbib,
      tightenlines,
      amsmath
]{revtex4-1}
\usepackage{bm}
\usepackage[dvips]{color,graphicx}
\usepackage{epsfig}
\usepackage{hyperref}
\hypersetup{colorlinks,%
            citecolor=black,%
            filecolor=black,%
            linkcolor=black,%
            urlcolor=black,%
            pdftex}

\def    \eg             {{\em e.g.}}
\newcommand{\ud}     {\mathrm{d}}
\newcommand{\gev}    {\ \mathrm{GeV}}
\newcommand{\gevsq}  {\ \mathrm{GeV}^2}
\newcommand{\mev}    {\ \mathrm{MeV}}
\newcommand{\ceps}{\varepsilon}
\newcommand{\eq}[1]{Eq.(\ref{#1})}
\newcommand{\Eqs}[2]{Eqs.~(\ref{#1}) and (\ref{#2})}

\begin{document}
\title{%
Structure Functions for Light Nuclei
}

\author{S.~A.~Kulagin}
\email[]{kulagin@ms2.inr.ac.ru}
\affiliation{Institute for Nuclear Research of the Russian Academy of Sciences, 117312 Moscow, Russia}
%
\author{R.~Petti}
\email[]{Roberto.Petti@cern.ch}
\affiliation{Department of Physics and Astronomy, University of South Carolina, Columbia SC 29208, USA }


\begin{abstract}
We discuss the nuclear EMC effect with particular emphasis on recent data 
for light nuclei including $^2$H, $^3$He, $^4$He, $^9$Be, $^{12}$C and $^{14}$N.
In order to verify the consistency of available data, we calculate the $\chi^2$ 
deviation between different data sets.  
We find a good agreement between the results from the NMC, SLAC E139, and HERMES experiments.
However, our analysis indicates an overall normalization offset of about $2\%$ in the data 
from the recent JLab E03-103 experiment with respect to previous data for nuclei heavier 
than $^3$He. We also discuss the extraction of the neutron/proton structure function ratio 
$F_2^n/F_2^p$ from the nuclear ratios $^3$He/$^2$H and $^2$H/$^1$H.  
Our analysis shows that the E03-103 data on $^3$He/$^2$H require a renormalization of about 
$3\%$ in order to be consistent with the $F_2^n/F_2^p$ ratio obtained from the NMC experiment.
After such a renormalization, the $^3$He data from the E03-103 data and HERMES experiments are in a good agreement.
Finally, we present a detailed comparison between data and model calculations,
which include a description of the nuclear binding, Fermi motion
and off-shell corrections to the structure functions of bound proton and neutron, 
as well as the nuclear pion and shadowing corrections.
Overall, a good agreement with the available data for all nuclei is obtained.  
\end{abstract}


\maketitle

\section{Introduction}
\label{sec:intro}

The nuclear EMC effect in the deep-inelastic scattering (DIS) has been
discussed since early 1980s starting from the observation of a dramatic
change in the structure function of a heavy nucleus relative to that of the deuteron
\cite{Aubert:1983xm}.
The nuclear effects in DIS were experimentally measured in the form of the 
ratio $\mathcal{R}(A/B)=F_2^A/F_2^B$ of the structure functions
(or the cross sections) of two nuclei
(usually a complex nucleus to deuterium) in the experiments
at CERN \cite{Bari:1985ga,Ashman:1992kv,Arneodo:1995cs,Amaudruz:1995tq,Arneodo:1996rv},
SLAC \cite{Gomez:1993ri}, DESY \cite{Ackerstaff:1999ac},
FNAL \cite{Adams:1995is}.
The data were taken for different nuclear targets and at
different regions of the Bjorken $x$ and the invariant momentum
transfer squared $Q^2$ and significant nuclear effects were observed
(for a review see \cite{Arneodo:1992wf,Geesaman:1995yd}).
Recently, the E03-103 experiment at Jlab published a high-statistics 
measurements of the EMC effect for ${}^3$He, ${}^4$He, ${}^9$Be, and ${}^{12}$C, nuclei
emphasizing the region of large $x$
\cite{Seely:2009gt}.
In this article we discuss the statistical consistency of data sets from different experiments 
collected with the same nuclear targets in the region $0.1<x<0.7$,
focusing on the kinematical region and the targets used in the E03-103 experiment.
We also study the sensitivity of the data on the EMC effect in ${}^3$He to the modeling of the ratio $F_2^n/F_2^p$.

A quantitative understanding the nuclear effects in DIS is important for a number of reasons.
A proper interpretation of experimental data can provide valuable insights into the origin of nuclear force
and helps us to understand how the properties of hadrons modify in a nuclear medium.
It should be also noted that the nuclear data often serve as the source of
information on hadrons otherwise not directly accessible. 
A typical example is the extraction of the neutron structure function which is
usually obtained from deuterium and proton data 
and requires a detailed knowledge of nuclear effects \cite{Kahn:2008nq}.
Other examples include the using of charged-lepton and neutrino nuclear data
in global QCD fits aiming to better determine the proton and neutron parton distribution functions
and the higher twist terms \cite{Alekhin:2009ye,Alekhin:2007fh,Alekhin:2008mb}.
Understanding the nuclear effects is particularly relevant
for precision measurements in neutrino physics, where heavy nuclear targets are used 
in order to collect a significant number of interactions.

A quantitative model for nuclear structure functions was recently developed in Refs.\cite{Kulagin:2004ie,Kulagin:2007ju}.
This approach accounts for a number of nuclear effects including nuclear Fermi motion 
and binding (FMB), nuclear pion excess, shadowing, and off-shell correction to bound nucleon 
structure functions.
The detailed analysis of data published before 1996 on the ratios $\mathcal{R}(A/B)$ for a wide region of nuclear targets
shows a good performance of the model which was able to describe the observed $x, Q^2$, and $A$ dependencies~\cite{Kulagin:2004ie}. 
In the present article we compare the predictions of our model with the recent data from HERMES~\cite{Ackerstaff:1999ac}
and E03-103~\cite{Seely:2009gt} experiments and discuss the role of different nuclear corrections in light nuclei.

The article is organized as follows.
In Sec.\ref{sec:model} we outline the model used in our studies.
In Sec.\ref{sec:discus} we discuss the data from different experiments and
also confront the results of our calculation with data.
In Sec.\ref{sec:sum} we summarize the results.

\section{Theoretical Model}
\label{sec:model}

We recall that while DIS is characterized by a large invariant momentum transfer $Q \gg M$, 
where $M$ is the nucleon mass, the characteristic longitudinal distance in the target rest frame
$L\sim 2q_0/Q^2=(Mx)^{-1}$ is not small in hadronic scale (see, \eg, Ref.\cite{Ioffe:1985ep}).
In nuclei, the comparison between $L$ and an average distance between bound nucleons $r_{NN}$ gives the
characteristic regions of the Bjorken variable $x$, which are goverened by different nuclear
effects.
As $r_{NN}\sim 1\div 2$\,fm, at large values of $x$ we have $L\ll r_{NN}$
and nuclear DIS can be approximated by the incoherent scattering off the bound protons and
neutrons (impulse approximation, or IA).
In this region the major nuclear corrections are due to nuclear binding 
and nucleon momentum distribution (Fermi motion).
At small values of $x$ the longitudinal distance $L$ becomes large and the IA is difficult to justify.
In this region the corrections due to scattering off nuclear pions (meson exchange currents),
as well as the nuclear (anti-)shadowing effect due to 
coherent multiple interactions of hadronic component
of virtual intermediate boson with bound nucleons, become important.
Summarizing, for the nuclear structure function we have different contributions 
(to be specific we discuss $F_2$, for more detail see Ref.\cite{Kulagin:2004ie}):
\begin{equation}
\label{F2A}
F_2^A = F_2^{\rm (IA)} + \delta_\pi F_2^A + \delta_{\rm coh} F_2^A,
\end{equation}
where the first term in the right-hand side stands for the impulse approximation,
and $\delta_\pi F_2$ and $\delta_{\rm coh} F_2$ are the corrections
due to scattering off the nuclear pion (meson) field and the
coherent interaction of the intermediate virtual boson with the nuclear target, respectively.

The IA term dominates at large $x$. This term can be written as a sum of the proton ($\tau=p$)
and the neutron ($\tau=n$) contributions \cite{Kulagin:2004ie}:
\begin{align}
\gamma^2 F_2^A(x,Q^2) &=
 \sum_{\tau=p,n}\int [\ud p]
	{\mathcal P}^\tau(\varepsilon,\bm{p})\,
		\left(1+\frac{\gamma p_z}{M}\right)
\left({\gamma'}^2 +\frac{6{x'}^2 \bm{p}_\perp^2}{Q^2} \right)
F_2^\tau(x',Q^2,p^2),
\label{IA}
\end{align}
where the integration is taken over the four-momentum of the bound nucleon
$p=(M+\ceps,\,\bm{p})$ and $[\ud p]=\ud\ceps\,\ud^3\bm{p}/(2\pi)^4$.
In the integrand, $\mathcal{P}^{\smash{p(n)}}(\varepsilon,\bm{p})$
is the proton (neutron) nuclear spectral function,
which describes the energy and momentum distribution of bound nucleons,
$F_2^{\smash{p(n)}}$ is the structure
function of bound proton (neutron), which
depends on the Bjorken variable $x'=Q^2/(2p q)$  
the momentum transfer
square $Q^2$ and also on the nucleon invariant mass squared $p^2=(M+\ceps)^2-\bm{p}^2$.
In \eq{IA} we use the coordinate system in which the momentum transfer $\bm{q}$
is antiparallel to the $z$ axis, $\bm{p}_\perp$ is the transverse component of the nucleon momentum,
and $\gamma^2=1+4x^2 M^2/Q^2$ and ${\gamma'}^2=1+4{x'}^2 p^2/Q^2$.

Nuclei typically have different proton and neutron numbers.
For this reason the nuclear structure functions generally have both the
isoscalar and the isovector contributions.
In order to separate the isoscalar and isovector contributions in
\eq{IA}, we introduce the isoscalar
$\mathcal{P}^{p+n}=\mathcal{P}^p+\mathcal{P}^n$
and the isovector $\mathcal{P}^{p-n}=\mathcal{P}^p-\mathcal{P}^n$
spectral functions. The spectral function $\mathcal{P}^{p+n}$ is normalized to the total nucleon number
$A=Z+N$ with $Z$ and $N$ being the proton and neutron number, respectively,
and $\mathcal{P}^{p-n}$ is normalized to  $Z-N$.
We separate the normalizations from the spectral functions and write
\begin{subequations}
\label{spfn:01}
\begin{align}
\label{spfn:0}
\mathcal{P}^{p+n} &= A \mathcal{P}_0,
\\
\mathcal{P}^{p-n} &= (Z-N)\mathcal{P}_1,
\label{spfn:1}
\end{align}
\end{subequations}
where the reduced spectral functions $\mathcal{P}_{0}$ and
$\mathcal{P}_{1}$ are both normalized to unity.
Using \eq{spfn:01} we explicitly write the nuclear structure function in \eq{IA} in
terms of the isoscalar and the isovector contributions
\begin{equation}
\label{nuke:FA}
F_2^A/A = \left\langle F_2^N \right\rangle_0 +
	\frac{\beta}{2} \left\langle F_2^{p-n} \right\rangle_1,
\end{equation}
where $F_2^N=\tfrac12(F_2^p+F_2^n)$ is the structure function of the isoscalar
nucleon and  $F_2^{p-n}=F_2^p-F_2^n$ 
and the parameter
$\beta=(Z-N)/A$ describes the fractional difference of protons and neutrons in a nucleus.
The quantities $\left\langle F \right\rangle_{0}$ and
$\left\langle F \right\rangle_{1}$ are the
contracted notations of the integration in
\eq{IA} taken with the reduced spectral
functions $\mathcal{P}_0$ and $\mathcal{P}_1$, respectively.
The model of $\mathcal{P}_0$ and $\mathcal{P}_1$, which is used in this article for nuclei with $A\ge4$,
is discussed in Sec.\ref{sec:spfn}. We also note that for $^3$He we apply \eq{IA} with the proton and neutron
spectral functions taken from microscopic calculations of Refs.\cite{SS,KPSV}.

It should be noted that the experimental data on the nuclear EMC effect have been often
presented for the isoscalar part of the nuclear structure functions
in order to facilitate comparison between different nuclei.
In order to separate the isoscalar part from nuclear data, 
a common practice is to multiply the data by a factor
\begin{equation}
\label{eq:iso}
C_{\mathrm is}=\frac{A F_2^N}{Z F_2^p+N F_2^n}=
\left(1-\beta \frac{F_2^{p-n}}{F_2^{p+n}} \right)^{-1} 
\end{equation}
that accounts for the proton-neutron difference in a nuclear target.
Indeed, neglecting nuclear effects we have from \eq{nuke:FA} $C_{\mathrm is} F_2^A=A F_2^N$.
It must be emphasized, however, that this method is only an approximate
way to isolate the isoscalar contribution to the structure functions, as a correct procedure
should involve the details of the proton and neutron distributions in nuclei.
Also, it is important to realize that the data are biased by this procedure as the factor 
$C_{\mathrm is}$ depends on the neutron structure function through the ratio $F_2^n/F_2^p$
and different experiments employ different models of this ratio.
For this reason the comparison between theoretical calculations and data is 
not clear-cut for nonisoscalar nuclei. Apparently, a consistent comparison between data 
and calculations for nonisoscalar nuclei should
involve unbiased data.
However, in practice it is not always possible to remove the isoscalar
correction from data of specific experiments.

\subsection{Nuclear Spectral Function}
\label{sec:spfn}

The nuclear spectral function $\mathcal{P}$ describes the energy and momentum 
distributions of bound nucleons and can be written as
\begin{equation}
\label{eq:spfn}
\mathcal{P}(\ceps,\bm{p}) = 2\pi
  \langle A|
	  a^{\dagger}(\bm{p})\delta(E_A-H-\ceps) a(\bm{p})
  | A\rangle ,
\end{equation}
where $a^{\dagger}(\bm p)$ and $a(\bm p)$ are creation and annihilation operators
of the nucleon with momentum $\bm p$, $E_A$ is the energy of the nuclear state $|A\rangle$,
and $H$ is the Hamiltonian of the system. Note that we discuss unpolarized scattering
and in \eq{eq:spfn} we implicitly assume the sum over nucleon polarizations. 
For simplicity, we also suppress explicit notations for the nucleon isospin state
(proton or neutron) of $\mathcal{P}$ and the operators $a^{\dagger}$ and $a$. 

Note also that \eq{eq:spfn} is written in the target rest frame and
defines the nuclear spectral function as a function of nucleon
energy $\ceps=p_0-M$ and momentum $\bm p$.
However, in the literature  the spectral function is usually considered
as a function of the nucleon separation energy $E$.
The spectral function in this case is denoted by $P(E,\bm p)$.
The relation between $\mathcal P(\ceps,\bm p)$ and $P(E,\bm p)$ is driven by a
relation between $\ceps$ and $E$, which  differ by the kinetic energy of a
recoil nuclear system of $A{-}1$ nucleons, $\ceps=-E-\bm{p}^2/2M_{A{-}1}$
(for more details see Ref.\cite{Kulagin:2008fm}).

A common way to calculate the spectral function is to insert a full set 
of the intermediate states in \eq{eq:spfn} and evaluate the sum of the 
corresponding nuclear matrix elements.
In the case of the deuteron, the intermediate states reduce to 
a free proton or neutron, and the spectral function is given entirely in terms of the
deuteron wave function square 
and $\ceps=E_D -\bm{p}^2/2M$, where $E_D\approx -2.22\mev$ the deuteron binding energy 
(the explicit form of \eq{IA} in this case is given in Ref.\cite{Kulagin:2004ie}).

For a $^3$He nucleus, the proton spectral function receives two contributions:
from the bound $(pn)$ intermediate state corresponding to a deuteron,
where the separation energy is $E = E_D - E_{^3{\rm He}}$, with
$E_{^3{\rm He}} \approx -7.72\mev$ the $^3$He binding energy; 
and from the $(pn)$ continuum scattering states. 
The neutron spectral function, on the other hand, has only the
$(pp)$ continuum contribution:
\begin{subequations}
\label{eq:spfn3He}
\begin{align}
\label{eq:pHe3}
P^p(E,\bm{p})
&= f^{p (d)}(\bm{p})\delta\left( E + E_{^3{\rm He}} - E_D \right)
 + f^{p ({\rm cont})}(E,\bm{p}),
\\ 
\label{eq:nHe3}
P^n(E,\bm{p})
&= f^{n ({\rm cont})}(E,\bm{p}) .
\end{align}
\end{subequations}

A number of calculations of $^3$He spectral function is available.
In Ref.\cite{SS} this spectral function was obtained by
solving the Faddeev equation with the Paris $NN$ potential \cite{paris:wf}
for the ground state wave function, and constructing its projection
onto the deuteron and two-body continuum states.
In Ref.\cite{KPSV} the spectral function was obtained using
a variational  $^3$He wave function calculated for a more recent AV14 $NN$ potential.
While in our numerical analysis we use the spectral functions from Ref.\cite{SS}, 
we verified that both Ref.\cite{SS} and Ref.\cite{KPSV} lead to a consistent
$\mathcal{R}$(${}^3$He/$^2$H) ratio.%
\footnote{%
In fact, the spectral functions of Ref.\cite{SS} and Ref.\cite{KPSV} result in an
almost identical EMC ratio for $x<0.85$. For $x>0.85$ and $Q^2$ values of the
E03-103 experiment, the calculation with the spectral function of Ref.\cite{KPSV}
gives $\mathcal{R}$(${}^3$He/$^2$H) larger by some 1--2\%.
For the spectral function of Ref.\cite{KPSV} we used a more recent results
with AV18 and Urbana IX potential (G.~Salme, private communication).
}

For $A\ge 4$ nuclei we follow the model of Ref.\cite{Kulagin:2004ie},
in which the full nuclear spectral function was calculated as a sum of the
mean-field spectral function, describing a low-energy part of the separation energy spectrum,
and the term generated by short-range $NN$ correlations in the nuclear ground state responsible
for a high separation energy and high-momentum component.

\begin{table}[htb]
\vspace{-2ex}
\begin{tabular}{l||c|c|c}\hline
Nucleus   & ~~$E_A/A$~~ & ~~~~$\langle\ceps\rangle$~~~~ & ~~$\langle\bm p^2\rangle/2M$~~  \\ \hline\hline
$^2$H       & $-1.11$     & $-11.46$                  & $9.24$   \\                    
${}^3$He    & $-2.57$     & $-17.95$                  & $12.87$  \\                   
${}^4$He    & $-7.07$     & $-40.06$                  & $25.01$  \\                   
${}^9$Be    & $-6.46$     & $-41.20$		      & $27.40$  \\ 
${}^{12}$C  & $-7.68$     & $-45.35$                  & $28.83$  \\                   
${}^{14}$N  & $-7.48$     & $-45.13$                  & $28.40$  \\
\hline
\end{tabular}
\caption{The nuclear binding energy per nucleon $E_A/A$,
a bound nucleon energy $\ceps$ 
and kinetic energy $\bm p^2/2M$  averaged with the nuclear spectral function
normalized to one nucleon (all in MeV units).
}
\label{tab:nuc}
\vspace{-1ex}
\end{table}

In order to illustrate the evolution of the nuclear binding effect in light nuclei,
in Table~\ref{tab:nuc} we list the values of nuclear binding energy along with the average
separation and kinetic energies of a bound nucleon for a few light nuclei.
For the deutron $^2$H we use the Paris wave function, the parameters for $^3$He
were calculated using the spectral function of Ref.\cite{SS}, whereas for $A\ge 4$  nuclei
a model spectral function of Ref.\cite{Kulagin:2004ie} was used.
Note a dramatic change in the energy parameters when going from 
$^3$He to $^4$He, which is also the underlying reason of a difference in 
the magnitude of nuclear corrections to the structure functions, as will be discussed 
in Sec.\ref{sec:discus}.

\subsection{Nucleon Structure Functions}
\label{sec:nsf}

In the DIS region, motivated by the twist expansion, the nucleon structure functions
can be written as
\begin{equation}\label{eq:SF}
F_2(x,Q^2) = F_2^{\text{TMC}}(x,Q^2)
        + H_2^{(4)}(x,Q^2)/Q^{2}
\end{equation}
where $F_2^{\text{TMC}}$ is the leading-twist (LT) structure function
corrected for the target mass effects and $H_2^{(4)}$ is the function describing the
contribution of the twist-four terms. The LT structure functions
are computed using the proton and neutron PDFs extracted from analysis of DIS and
Drell-Yan data in a kinematical region of $Q^2>1\gevsq$ and invariant mass $W>1.8\gev$ 
\cite{Alekhin:2002fv,Alekhin:2006zm} with the coefficient functions
calculated to the NNLO approximation 
in the strong coupling constant in pQCD \cite{Zijlstra:1992qd}. The target mass
correction is computed following Ref.\cite{Georgi:1976ve}.
%
Although in general the twist expansion should include an infinite chain of the HT power terms,
recent phenomenology suggests that \eq{eq:SF} with only twist-four correction
provides a good description of data down to $Q\sim 1\gev$
\cite{Alekhin:2002fv,Alekhin:2003qq,Alekhin:2006zm,Alekhin:2007fh}.
It is also worth noting that this model is consistent with the duality principle and on average
describes the resonance data with $W<1.8\gev$ \cite{Alekhin:2007fh}, which is
relevant for the kinematical region of the JLab E03-103 experiment \cite{Seely:2009gt}.

In calculating the nuclear structure functions one has to deal with 
the structure functions of the bound proton and neutron, which generally
differ from those of the free proton and neutron. 
The effect of modification of bound nucleon SF
is related to 
the dependence on the nucleon invariant mass $p^2$ in the off-shell region
[see \eq{IA}].
One can separate the two sources of $p^2$ dependence: 
(i) The $p^2$ terms originating from the target mass effect in the off-shell region.
The corresponding correction is of the order $p^2/Q^2$.
In Ref.\cite{Kulagin:2004ie} this effect is evaluated by the
replacement $M^2\to p^2$ in the expressions of Ref.\cite{Georgi:1976ve}.
(ii) The off-shell dependence of the the parton distributions (or LT structure functions);
the latter effect is generally not suppressed by the inverse powers of $Q^2$.
In the description of this effect we treat the nucleon virtuality $v=(p^2-M^2)/M^2$ as a small parameter, expand in series in $v$ keeping the leading term:
\begin{align}
\label{SF:OS}
F_2^{\mathrm{LT}}(x,Q^2,p^2) &=
F_2^{\mathrm{LT}}(x,Q^2)\left( 1+\delta f_2(x,Q^2)\,v \right),\\
\label{delf}
\delta f_2 &= \partial\ln F_2^{\mathrm{LT}}/\partial\ln p^2,
\end{align}
where the first term on the right in \eq{SF:OS} is the structure
function of the on-mass-shell nucleon and the derivative is evaluated
at $p^2=M^2$. 

Although the off-shell function (\ref{delf}) generally
depends on the structure function type,
it was suggested in Refs.\cite{Kulagin:2004ie,Kulagin:2007ju} that
the relative off-shell effect (\ref{delf}) is common for all types of
the nucleon PDFs. Thus, the function $\delta f_2$ measures a relative
response of the nucleon parton distributions to the variation of the nucleon
mass and, after the averaging with the nuclear spectral function,
it describes the modification of the nucleon PDFs in the nuclear environment.

\section{Comparison with Data}
\label{sec:discus}

We start this section by summarizing the results of our previous analysis~\cite{Kulagin:2004ie} 
of data on the ratios of structure functions $\mathcal{R}(A/B)=F_2^A/F_2^B$ published before 1996.   
In Ref.\cite{Kulagin:2004ie} we performed a statistical analysis by calculating  
$\chi^2=\sum (\mathcal{R}^\mathrm{exp}-\mathcal{R}^\mathrm{th})^2/\sigma^2(\mathcal{R}^\mathrm{exp})$,
where $\mathcal{R}^\mathrm{exp}$ and $\mathcal{R}^\mathrm{th}$ are the data points and model predictions,
respectively, and  $\sigma$ is the uncertainty of the experimental data points.
The examined data come from different experiments for a variety of targets
ranging from $^4$He to $^{208}$Pb with $Q^2\ge 1\gevsq$ and in a wide region of Bjorken $x$ 
(for a complete list of the analyzed data see Table~1 of Ref.\cite{Kulagin:2004ie}).

\begin{table}[!]
\vspace{-2ex} 
\begin{tabular}{l||c|c|c|c|c|c|c} \hline  
Targets~~~~ & \multicolumn{7}{c}{$\chi^2/$DOF} \\  
 &  ~~~NMC~~~ &  ~~~EMC~~~ & ~~~E139~~~ & ~~~E140~~~ & ~~~BCDMS~~~ & ~~~E665~~~ & ~~~HERMES~~~ \\ \hline\hline  
 ${}^4$He/$^2$H &  10.8/17 &   & 6.2/21 &  &  &  &  \\   
 ${}^7{\rm Li}$/$^2$H & 28.6/17  &   &  &  &  &  &  \\   
 ${}^9{\rm Be}$/$^2$H &   &   & 12.3/21  &  &  &  &  \\   
 ${}^{12}$C/$^2$H &  14.6/17 &   & 13.0/17 &  &  &  &  \\   
 ${}^{9}{\rm Be}/{}^{12}$C & 5.3/15  &   &  &  &  &  &  \\   
 ${}^{12}$C/${}^7{\rm Li}$ & 41.0/24  &   &  &  &  &  &  \\   
 ${}^{14}$N/$^2$H &   &   &  &  &  &  & 9.8/12 \\   
 ${}^{27}$Al/$^2$H &   &   & 14.8/21 &  &  &  &  \\   
 ${}^{27}$Al/C & 5.7/15  &   &  &  &  &  &  \\   
 ${}^{40}$Ca/$^2$H & 27.2/16  &   &  14.3/17 &  &  &  &  \\   
 ${}^{40}$Ca/${}^7{\rm Li}$ & 35.6/24  &   &  &  &  &  &  \\   
 ${}^{40}$Ca/${}^{12}$C & 31.8/24  &   &  &  &  & 1.0/5  &  \\   
 ${}^{56}{\rm Fe}$/$^2$H &   &   & 18.4/23 & 4.5/8 & 14.8/10 &  &  \\   
 ${}^{56}{\rm Fe}$/${}^{12}$C & 10.3/15  &   &  &  &  &  &  \\   
 ${}^{63}{\rm Cu}$/$^2$H &   & 7.8/10  &  &  &  &  &  \\   
 ${}^{84}$Kr/$^2$H   &   &   &  &  &  &  & 4.9/12 \\   
 ${}^{108}{\rm Ag}$/$^2$H &   &   & 14.9/17 &  &  &  &  \\   
 ${}^{119}{\rm Sn}$/${}^{12}$C & 94.9/161  &   &  &  &  &  &  \\   
 ${}^{197}{\rm Au}$/$^2$H &   &   & 18.2/21 & 2.4/1 &  &  &  \\   
 ${}^{207}{\rm Pb}$/$^2$H &   &   &  &  &  & 5.0/5 &  \\   
 ${}^{207}{\rm Pb}$/${}^{12}{\rm C}~$ & 6.1/15  &   &  &  &  & 0.2/5  &  \\ \hline    
\end{tabular}
\caption{%
Values of $\chi^2/$DOF between different data sets with $Q^2\geq 1$ GeV$^2$ and
the predictions of Ref.\cite{Kulagin:2004ie}. The normalization of each experiment is fixed.
The sum over all data results in $\chi^2{\rm /DOF}=466.6/586$.   
}
\vspace{-2ex}
\label{tab:chisq-all}   
\end{table}

The ratio $\mathcal{R}^\mathrm{th}$ was calculated using approach  outlined in Sec.\ref{sec:model}.
We note that the calculations with \eq{IA} with the realistic nuclear spectral function, accounting
for the Fermi motion and nuclear binding effect (FMB), and no off-shell correction explains a general
trend of the EMC effect at large $x$ values
\cite{Akulinichev:1985ij,Akulinichev:1985xq,Kulagin:1989mu}.
Nevertheless, the FMB effects alone fail to quantitatively describe data.
In Ref.\cite{Kulagin:2004ie} a hypothesis was tested that the account of the off-shell effect
in the bound nucleon SF would provide an accurate  description of data.
The function $\delta f_2$ was assumed to depend only on $x$ which was then treated phenomenologically
as a polynomial parametrization. Then the parameters, together 
with their uncertainties, were extracted from statistical analysis of data.
This approach lead to an excellent agreement with all available data on $\mathcal{R}(A/B)$,  
reproducing the observed $x$, $Q^2$, and $A$ dependencies. 
In order to demonstrate the consistency between our model and the data,
in Table~\ref{tab:chisq-all} we list the values of $\chi^2{\rm /DOF}$
obtained from data on different nuclei and from different experiments.

\subsection{Data from nuclear targets with $A\geq 4$}

\begin{figure}[p]
\begin{center}
\vspace{-2cm}\epsfig{file=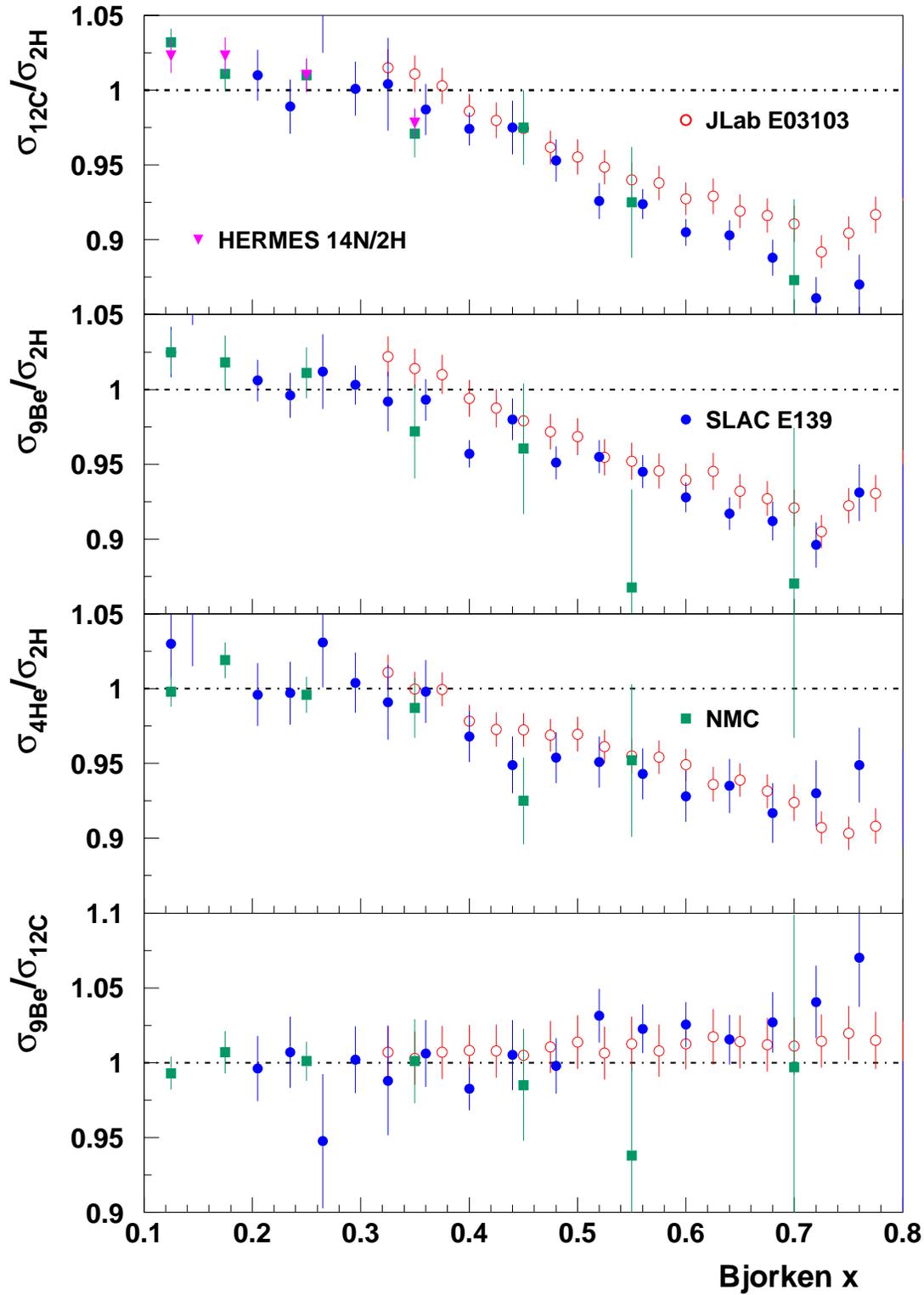,width=0.95\textwidth}
\caption{(Color online)
The $\mathcal{R}$-ratios of ${}^{14}$N, ${}^{12}$C, ${}^9$Be, ${}^4$He with respect to deuterium 
and the ratio ${}^9$Be/${}^{12}$C 
as measured by the NMC~\cite{Arneodo:1995cs,Amaudruz:1995tq,Arneodo:1996rv} (full squares),
SLAC E139~\cite{Gomez:1993ri} (full circles), JLab E03-103~\cite{Seely:2009gt} (open circles) 
and HERMES~\cite{Ackerstaff:1999ac} (full triangles) experiments.   
Statistical and systematic uncertainties are added in quadrature, while the
normalization uncertainty is not shown. For the E139 and E03-103 experiments the ratio ${}^9$Be/${}^{12}$C 
is calculated as a double ratio of data on ${}^9$Be/${}^2$H and ${}^{12}$C/${}^2$H.  
}
\label{fig:rdat}
\end{center}
\end{figure}

More recent data on the nuclear EMC effect are available from 
HERMES experiment at HERA~\cite{Ackerstaff:1999ac} and E03-103 experiment at JLab~\cite{Seely:2009gt}.
The HERMES experiment published the data for $^3$He/$^2$H, $^{14}$N/$^2$H and $^{84}$Kr/$^2$H ratios
obtained with a $27.5 \gev$ positron beam
for Bjorken $x$ in the range $0.0125<x<0.35$ and an average $Q^2$ ranging from about $0.5\gevsq$
in a low $x$ region to about $4\gevsq$ at large $x$.
The E03-103 experiment published the measurements of 
$^3$He/$^2$H, $^4$He/$^2$H, $^9$Be/$^2$H, and $^{12}$C/$^2$H ratios
with a $5.77 \gev$ electron beam for $x$ values ranging from $0.325$ to $0.95$ and 
an average $Q^2$ from about $3 \gevsq$ at $x\sim 0.3$ to $6\gevsq$ at high $x$.
In Fig.\ref{fig:rdat} we summarize these data together with the former data 
by the NMC and SLAC E139 collaborations.  


\begin{table}[!] 
\vspace{-5ex}
\begin{tabular}{l||cccc} \hline  
Targets~~~~  & ~~~~NMC~~~ & ~~~SLAC E139~~~ & ~~~HERMES~~~~ & ~~~~JLab E03-103~~~~ \\  \hline\hline  
${}^3$He/$^2$H   &            &                 &    1.4\%      &  1.84\%  \\ 
${}^4$He/$^2$H   &  0.4\%     & 2.4\%           &               &  1.5\%  \\ 
${}^9$Be/$^2$H   &  0.4\%     & 1.2\%           &               &  1.7\%  \\ 
${}^{12}$C/$^2$H &  0.4\%     & 1.2\%   &  1.4\% (${}^{14}$N/$^2$H) &  1.6\%  \\ \hline 
\end{tabular} 
\caption{Overall normalization uncertainties on the ratios $\cal{R}$ from different experiments.} 
\label{tab:norm} 
\bigskip  
\begin{tabular}{l||ccc}\hline   
 Experiment~~~~ & ~~~~SLAC E139~~~~ &  ~~~~NMC~~~~ & ~~~~HERMES~~~~ \\ \hline\hline  
 & \multicolumn{3}{c}{$\chi^2$/DOF for ${}^{12}$C/$^2$H:} \\  
SLAC E139 &  &  0.7/3  & 0.2/2  \\  
NMC &  &  & 0.1/3  \\ \hline   
 & \multicolumn{3}{c}{$\chi^2$/DOF for ${}^9$Be/$^2$H:} \\  
SLAC E139 &  &  0.7/3  &   \\ \hline  
 & \multicolumn{3}{c}{$\chi^2$/DOF for ${}^4$He/$^2$H:} \\  
SLAC E139 &  &  2.2/3  &   \\ \hline  
\end{tabular} 
\caption{%
Compatibility between SLAC E139, NMC and HERMES data in the region $0.1<x<0.5$. 
The data shown in Fig.\ref{fig:rdat} have been rebinned with a bin size of 
0.1 for all experiments. The normalization of each experiment is fixed. 
The normalization uncertainties are not included in the evaluation of $\chi^2$.  
}  
\label{tab:SLAC-norm}
\bigskip  
\begin{tabular}{l||ccc|c}\hline   
  \multicolumn{5}{c}{$\chi^2$/DOF between SLAC E139 and JLab E03-103} \\  
  & ${}^4$He/$^2$H  &  ${}^9$Be/$^2$H & ${}^{12}$C/$^2$H & Total \\ \hline\hline  
Default normalization & ~~~~4.6/4~~~~  & ~~~~21.7/4~~~~  & ~~~~16.4 /4~~~~  &  ~~~~42.7/12~~~~  \\  
Normalization factor 0.98~~~ &  1.1/4  & 6.4/4 &  1.3/4  & 8.8/12 \\ \hline   
\end{tabular} 
\caption{%
Compatibility between SLAC E139 and JLab E03-103 data in the region $0.3<x<0.7$. 
The data shown in Figure~\ref{fig:rdat} have been rebinned with a bin size of 0.1 for all experiments. In the second row all JLab data have been rescaled by an overall factor 0.98.
The normalization uncertainties are not included in the evaluation of $\chi^2$.   
}  
\label{tab:JLab-norm}  
\bigskip
\begin{tabular}{l||ccc|c} \hline  
Targets~~~~ & ~~~~$\chi^2/$DOF~~~ & ~~~Normalization~~~ & ~~~Normalization~~~~ & ~~~~$\chi^2/$DOF~~~~ \\  
 &  &  factor & uncertainty & ~~~~w/o offset~~~~ \\ \hline\hline  
${}^4$He/$^2$H  &  11.9/20 & $1.9\pm0.3\%$ & 1.50\% & 23.6/21 \\   
${}^9$Be/$^2$H  &  9.2/20 & $2.0^{+0.1}_{-0.5}\%$ & 1.70\% & 23.0/21 \\   
${}^{12}$C/$^2$H  & 5.2/20 & $2.1^{+0.2}_{-0.5}\%$ & 1.60\% & 27.0/21 \\  \hline  
All data  &  26.3/60 &&& 73.6/63 \\ \hline  
\end{tabular}
\caption{Values of $\chi^2/$DOF between JLab E03-103 data~\cite{Seely:2009gt} with 
$W^2>2$ GeV$^2$ ($x<0.85$) and our predictions~\cite{Kulagin:2004ie}. 
The overall normalization of each data set has 
been kept floating while the corresponding normalization uncertainty is not included 
in the $\chi^2$. For comparison, the last column shows the $\chi^2/$DOF 
obtained without any normalization offset but including the normalization uncertainty 
to the evaluation of $\chi^2$.
}
\label{tab:chisq-JLab}
\end{table}

\subsubsection{Consistency of different data sets}
\label{sec:consistency}

In this section we study the consistency of data from different experiments.  
From Fig.\ref{fig:rdat} we can conclude that the slopes of the EMC ratio as a function   
of $x$ from different experiments are in a good agreement. Note that
the beam energy of these experiments differs significantly.
As the beam energy determines the range of possible $Q^2$ values for each $x$-bin,
the typical $Q^2$ values of the NMC experiment are significantly
higher than those of the SLAC E139, HERMES, and JLab E03-103 experiments. 
Nevertheless, the data show no systematic $Q^2$ dependence of $\mathcal{R}$ in the 
valence region $0.25 < x < 0.6$. 


Contrary to the slope, the overall normalization factors of different experiments 
are not fully consistent.
If we consider the $^{12}$C and $^{14}$N nuclei, which have a similar atomic number,
we observe a good agreement in the normalizations of the NMC, SLAC E139, and HERMES 
experiments.
However, the data points of the JLab E03-103 experiment are somewhat shifted above the data 
from other experiments. 
A similar trend is also present for the other nuclei, as can be seen from Fig.\ref{fig:rdat}.
It is interesting to observe that for the double ratio (${}^9$Be/$^2$H)/(${}^{12}$C/$^2$H) the normalization 
of the E03-103 data seems to agree with the SLAC E139 and NMC data. This fact points 
toward a common normalization offset for all ratios, possibly related to a common 
source of systematic uncertainty.   
It should be noted that the overall normalization uncertainty is not shown in 
Fig.\ref{fig:rdat} but is summarized in Table~\ref{tab:norm}.   
The NMC experiment achieved the most precise absolute normalization within 0.4\%, 
followed by the SLAC E139 experiment. For this reason we try to use NMC data as a reference and 
we evaluate the consistency of other experiments with NMC. From Fig.\ref{fig:rdat} 
we can see that the statistical uncertainties of NMC data become large at $x>0.3$ and 
therefore we can have an accurate comparison only in the region $0.1<x<0.3$. Following this 
observation, we proceed in two steps. First, we study the consistency of SLAC E139 and 
HERMES data with NMC in this overlap region. Then we compare the E03-103 data with 
SLAC E139 data since those two experiments have a broader overlap region for $x>0.3$.

In order to evaluate the compatibility in the overall normalization factors of different 
experiments we follow a $\chi^2$ approach. We consider the data shown in Fig.\ref{fig:rdat} 
in the region $0.1<x<0.7$ and we rebin the measurements with a common bin size of 0.1/bin 
(six bins in total from 0.1 to 0.7). Within each larger bin, we calculate the weighted average 
of the data points from each experiments with weights given by $1/\sigma^2$. For each pair 
of experiments, we then calculate the corresponding $\chi^2$ in the overlap region assuming 
a fixed normalization, as reported by the experiments. The results are summarized in 
Tables~\ref{tab:SLAC-norm} and~\ref{tab:JLab-norm}. From Table~\ref{tab:SLAC-norm} we can 
conclude that the three 
experiments NMC, SLAC E139 and HERMES appear to be statistically compatible in the 
absolute normalization.  
However, the picture differs substantially  for the E03-103 experiment. The JLab kinematics 
largely overlaps with the SLAC E139 experiment and for this reason  we calculate the 
overall $\chi^2$ between these two measurements.
The results shown in Table~\ref{tab:JLab-norm} signal on inconsistency in the normalization
of data from these experiments.
We found that an overall normalization factor of $0.98^{+0.005}_{-0.003}$,
applied to the ${}^4$He/$^2$H, ${}^9$Be/$^2$H, and ${}^{12}$C/$^2$H data of the E03-103 experiment,
makes them statistically compatible with the 
SLAC E139 data, lowering $\chi^2$ by $\Delta \chi^2 = 33.9$. Therefore, in the following comparisons 
we will apply an overall normalization factor of 0.98 to the measurements 
${}^4$He/$^2$H, ${}^9$Be/$^2$H and ${}^{12}$C/$^2$H from Ref.~\cite{Seely:2009gt}.  

\subsubsection{Comparison with model predictions}
\label{sec:comparison}

\begin{figure}[!] 
\begin{center}
\vspace{-3ex}\epsfig{file=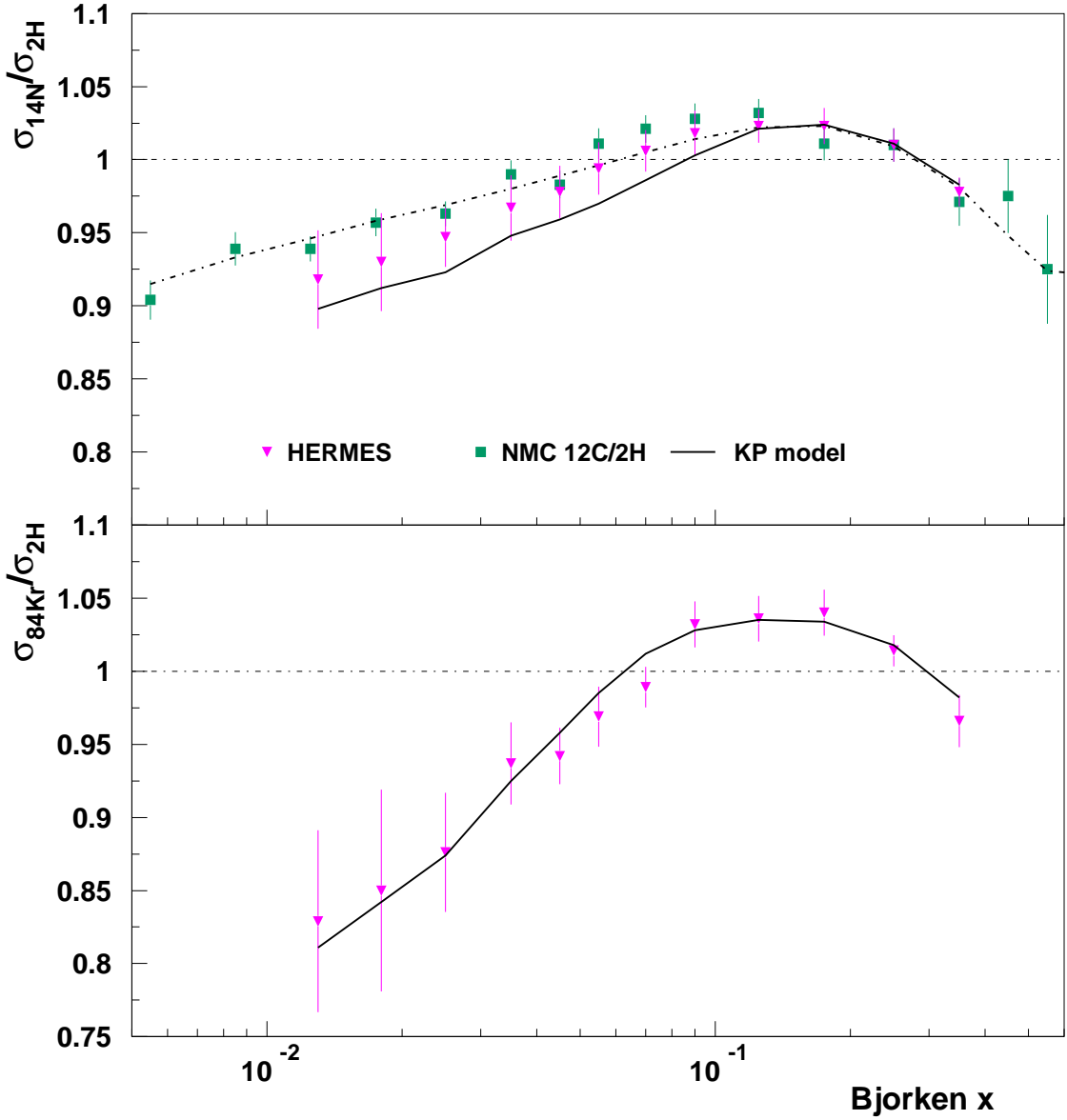,width=\textwidth}
\caption{(Color online)
Ratios of cross sections for positron scattering off ${}^{14}$N and 
${}^{84}$Kr nuclear targets with respect to a deuterium target. 
Data from the HERMES experiment~\cite{Ackerstaff:1999ac} (full triangles)
are compared with predictions of Ref.\cite{Kulagin:2004ie} calculated 
for the same kinematics (solid lines). Statistical and systematic uncertainties 
are added in quadrature, while the normalization uncertainty is not shown. 
The data from the NMC experiment for ${}^{12}$C target and the results of Ref.\cite{Kulagin:2004ie}
(dash-dotted line) are also shown in the top 
panel to illustrate the effect of the $Q^2$ dependence at low $x$ values (see text).   
}
\label{fig:hermes}
\end{center}
\end{figure}

Figure~\ref{fig:hermes} shows a comparison of our calculations with HERMES data \cite{Ackerstaff:1999ac}
and also  $^{12}$C/$^2$H data from NMC experiment.
These data emphasize the region of small $x$ where the major nuclear correction is due to nuclear shadowing. 
We found a good agreement of model calculations with data over the whole kinematical region of $x$ and $Q^2$. 
We note that the shadowing effect arise in our model due to medium effects in the propagation
of hadronic component of the virtual photon in nuclear environment.
The magnitude of the shadowing effect is driven by the total cross section of scattering of hadronic component
off a bound nucleon. This quantity strongly depends on $Q^2$ in the region of low $Q^2$~\cite{Kulagin:2004ie},
which is the underlying reason for the difference in the shadowing correction
for the NMC and HERMES data in Fig.\ref{fig:hermes}.
We note that characteristic $Q^2$ are significantly smaller for the HERMES experiment.
For example, in the $x$-bin $0.018$ the average $Q^2$ is $2.9\gevsq$ for NMC 
and about $0.66\gevsq$ for HERMES. It should be also noted 
that pQCD and the twist expansion approach employed in \eq{eq:SF} is out of the region of applicability
for such low values of $Q^2$. 
In order to evaluate the structure functions and their ratios at low $Q^2$, we apply the spline
extrapolation of \eq{eq:SF} from the region $Q^2\ge 1\gevsq$ to $Q^2\to 0$ using the requirements
from the conservation of the electromagnetic current \cite{Kulagin:2004ie}.
We also note that a bump in the EMC ratio between $x=0.1$ and 0.3 is due to an interplay between the
nuclear shadowing and pion correction effects.


\begin{figure}[p]
\begin{center}
\vspace{-2cm}\epsfig{file=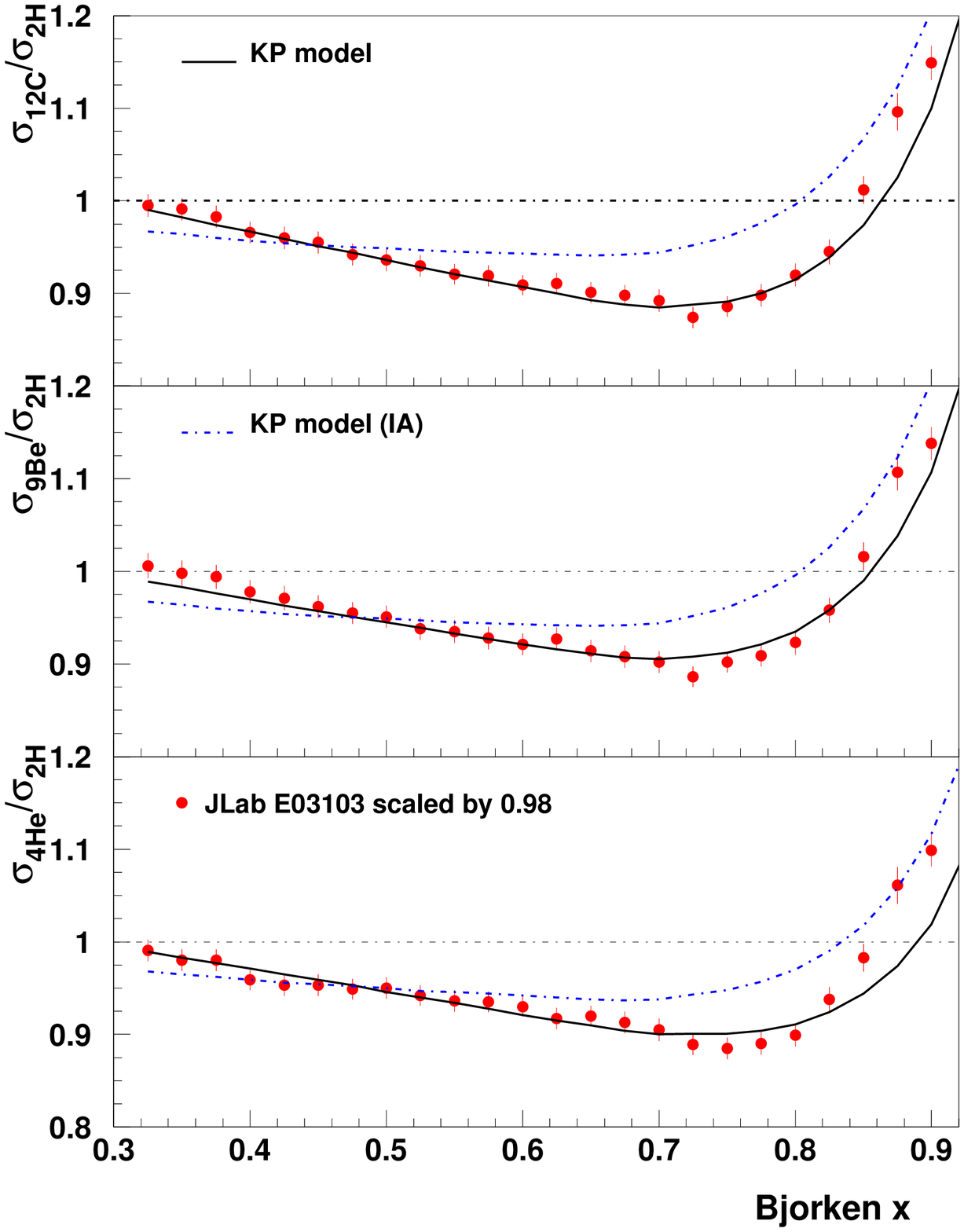,width=\textwidth}
\caption{(Color online)
Data on the $\mathcal{R}$ ratios of ${}^{12}$C, ${}^9$Be, and ${}^4$He with respect to 
deuterium compared 
with predictions of the model of Ref.\cite{Kulagin:2004ie} for the same kinematics.
A common normalization factor of $0.98$ has been applied to all data points of Ref.\cite{Seely:2009gt},
and statistical and systematic uncertainties are added in quadrature.
The result of a calculation in impulse approximation with no off-shell correction
is also shown as dashed-dotted line for comparison.
}
\label{fig:e03103}
\end{center}
\end{figure}

In Fig.\ref{fig:e03103} we show a detailed comparison of our predictions with the E03-103 data for the ratios
$^4$He/$^2$H, $^9$Be/$^2$H, and $^{12}$C/$^2$H. For each $x$ bin, the ratios of nuclear cross sections were 
calculated at the average $Q^2$ value provided by the E03-103 experiment~\cite{Seely:2009gt}.
After applying the normalization factor of 0.98, required to match the NMC, E139 and HERMES data, 
we observe a very good agreement between our model and the E03-103 experiment for $x<0.85$.
Table~\ref{tab:chisq-JLab} lists the results of calculation of $\chi^2$
between the model predictions and the data with and without renormalization.
As a cross check, we also list the values of a normalization factor minimizing
$\chi^2$ if we leave the normalization floating for each of the three ratios 
$^4$He/$^2$H, $^9$Be/$^2$H, and $^{12}$C/$^2$H from the E03-103 experiment.
Note that the 2\% offset appears
statistically compatible with the overall normalization uncertainties quoted by the 
E03-103 experiment. Indeed, if we keep fixed the JLab normalization and we rather add the 
corresponding normalization uncertainties into the $\chi^2$ calculation, we obtain 
a normalized $\chi^2$ of 1.17.
We also remark that allowing the normalizations to float reduces the $\chi^2$/DOF to
significantly less than unity that may signal an issue with the quoted uncertainties on the data.

In the region of $x>0.85$ our calculations somewhat underpredict the E03-103 data. 
We note that in this region the data of Ref.\cite{Seely:2009gt} have $W^2 < 2\gevsq$ and 
therefore lie within the resonance region, while in our calculations we use a DIS 
model for the structure functions.
Although we verified the consistency of low-$Q$ DIS structure functions with the duality 
principle, a more detailed comparison would require explicit resonance model for the 
structure functions.
Furthermore, the region of $x>0.85$ is very sensitive to the 
treatment of both the bound nucleon momentum distribution 
and the target mass correction to the nucleon structure functions~\cite{Kulagin:2004ie}. 
In our model, we apply the target mass correction following a commonly used method of  Ref.\cite{Georgi:1976ve}.
However, it is known that the target mass correction of Ref.\cite{Georgi:1976ve} leads to 
an incorrect behavior of SF in the limit of $x \to 1$ that has a significant 
impact on the calculation of nuclear ratio $\cal{R}$ at very large $x$ values.   
We also note that in the detailed comparison with the data the quasielastic contribution,
which is not addressed in this article, may not be negligible in this region.

Finally, Fig.\ref{fig:e03103} shows the results obtained with and without ($\delta f_2=0$) 
off-shell correction.
We recall that the off-shell function $\delta f_2$ was determined in the former analysis of 
data \cite{Kulagin:2004ie}.
We conclude that the off-shell correction plays a crucial role in understanding 
the data, implying the EMC effect is largely driven by the modification of  
properties of the bound nucleons inside the nuclei.

\subsection{Data from $^3$He target and the ratio $F_2^n/F_2^p$}

Here we discuss nuclear effects in the $^3$He target and futher address the problem of
consistency between data from different experiments. 
The EMC effect in the $^3$He nucleus was measured by HERMES~\cite{Ackerstaff:1999ac} and recently at JLab~\cite{Seely:2009gt}.
We recall that the data are presented in a form corrected for the proton--neutron difference
in a nucleus, as discussed in Sec.\ref{sec:model}. The isoscalar correction depends on the 
ratio $F_2^n/F_2^p$ and for this reason a direct comparison of corrected data is biased by the models
used in different experiments. 
The uncorrected data on the EMC ratios are available from Ref.\cite{Seely:2009gt}
and comparison with those data allows us to reduce a bias associated with the isoscalar correction.

In the former discussion we addressed  possible inconsistency in the normalization
of the EMC ratios from different experiments. Below we discuss the relation between the ratios
$\mathcal R({}^3\mathrm{He}/{}^2\mathrm{H}) = \tfrac13 F_2^{3\mathrm{He}}/\tfrac12 F_2^D$
and
$\mathcal R({}^2\mathrm{H}/{}^1\mathrm{H}) = \tfrac12 F_2^D/F_2^p$ and $F_2^n/F_2^p$,
and will argue that the comparison of the $F_2^n/F_2^p$ ratio extracted
from ${}^3$He/$^2$H data of the E03-103 experiment~\cite{Seely:2009gt} 
and the $^2$H/$^1$H data by NMC~\cite{Arneodo:1997dp} provides a sensitive test of 
the normalization of the  ${}^3$He/$^2$H data.

In the absense of nuclear effects, the ratio $F_2^n/F_2^p$  can directly be calculated from
either the $^2$H/$^1$H  or  $^3$He/$^2$H ratio. A realistic analysis should include
the treatment of nuclear effects.
Let us first consider the ratio $^2$H/$^1$H.
In order to address the nuclear corrections in this problem, it is convenient to consider
the ratio $R_2=F_2^D/(F_2^p+F_2^n)$ and recast $F_2^D$ in terms of $R_2$. We have for $F_2^n/F_2^p$:
\begin{equation}\label{eq:n_p:d}
F_2^n/F_2^p = 2\mathcal R({}^2\mathrm{H}/{}^1\mathrm{H})/R_2 - 1.
\end{equation}
Addressing the ratio $^3$He/$^2$H, we need to consider
$R_3=F_2^{3\mathrm{He}}/(2F_2^p+F_2^n)$ along with $R_2$.
We recast $F_2^{3\mathrm{He}}$ in terms of $R_3$ and after a simple algebra we have:
\begin{equation}\label{eq:n_p:he3}
F_2^n/F_2^p = (2-z)/(z-1),
\end{equation}
where $z=\tfrac32 \mathcal R({}^3\mathrm{He}/{}^2\mathrm{H}) R_2/R_3$.  

\begin{figure}[p]
\begin{center}
\vspace{-4ex}\epsfig{file=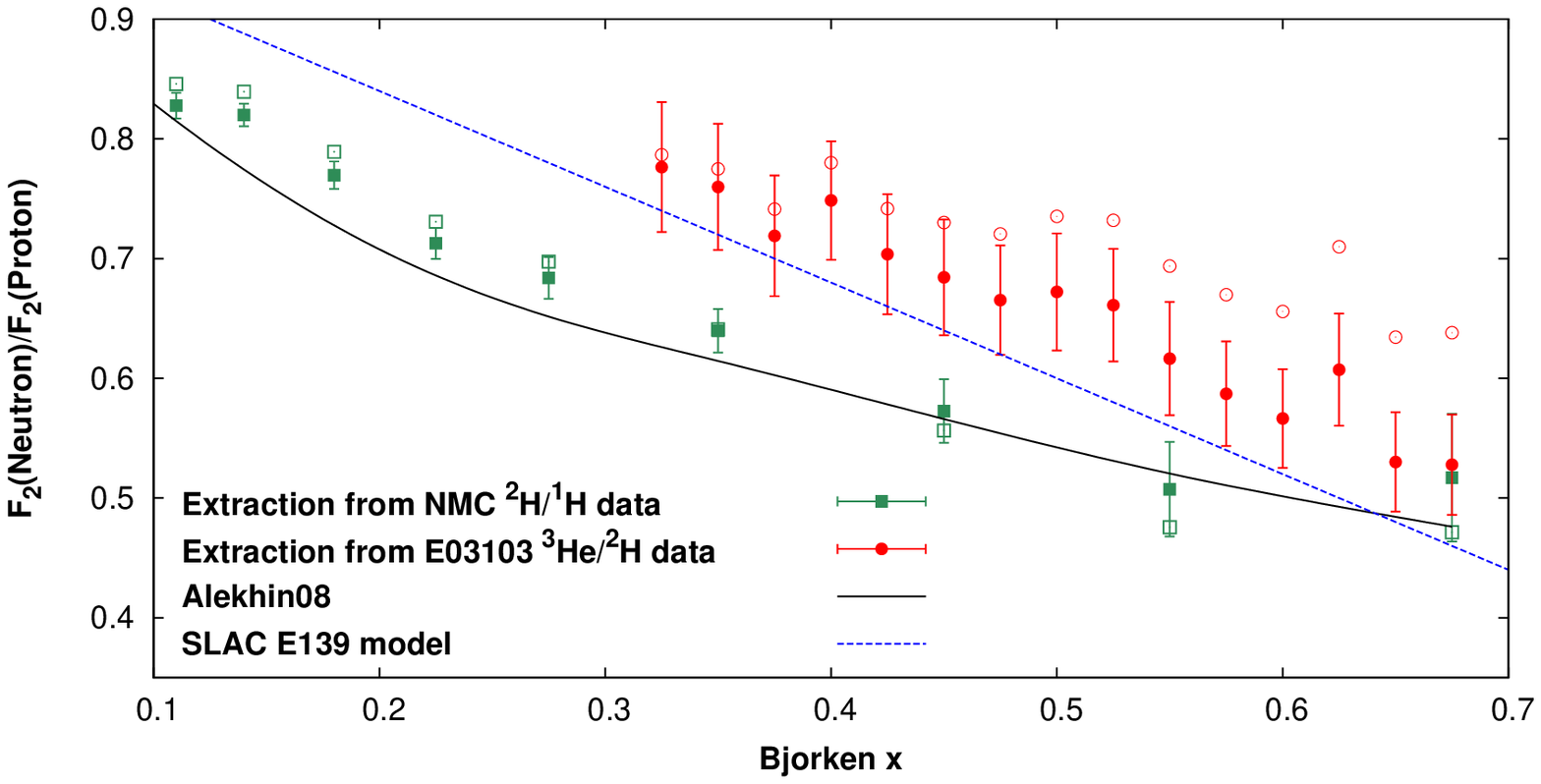,width=\textwidth}
\caption{(Color online)  
The results of the extraction of the ratio $F_2^n/F_2^p$ from data on the $^2$H/$^1$H cross-section ratio 
by the NMC experiment~\cite{Arneodo:1997dp} (green squares)
and the data on the $^3$He/$^2$H cross-section ratio 
by JLab E03-103 experiment~\cite{Seely:2009gt} (red circles).
The full symbols are obtained with the model of nuclear effects of Ref.\cite{Kulagin:2004ie},  
while the open symbols show the points obtained with no nuclear corrections.
The solid curve is the result of a calculation according to \eq{eq:SF}    
with the NNLO PDFs and the HT corrections of 
Refs.~\cite{Alekhin:2006zm,Alekhin:2007fh,Alekhin:2008ua}, for the $Q^2$ corresponding to 
the NMC kinematics.
The dashed line is a $Q^2$ independent model used in Ref.\cite{Gomez:1993ri}.
\label{fig:npratio}
}
\bigskip
\epsfig{file=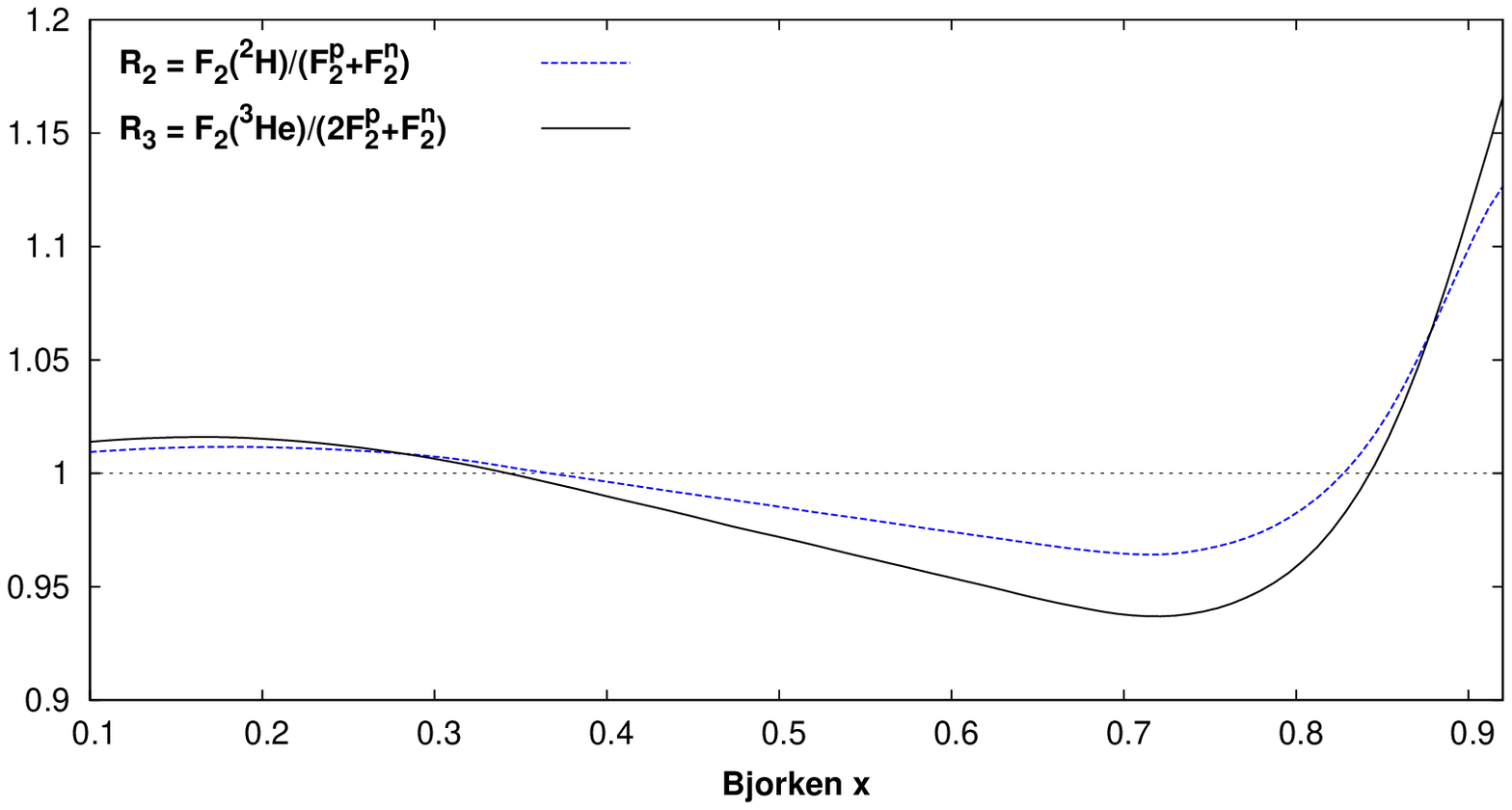,width=\textwidth}
\caption{(Color online)
\protect
The ratios $R_2$ and $R_3$ calculated at the values of $x$ and $Q^2$ of Ref.\cite{Seely:2009gt}
for $x>0.3$ and at fixed $Q^2=3\gevsq$ for $x<0.3$.
\label{fig:r2_r3}
}
\end{center}
\end{figure}

In Fig.~\ref{fig:npratio} we show the results of extraction of $F_2^n/F_2^p$  from the NMC
and the E03-103 data using \Eqs{eq:n_p:d}{eq:n_p:he3}.
It should be remarked that a typical region of $Q^2$ is significantly higher in the
NMC experiment than that of the E03-103 experiment at JLab. Nevertheless,
Ref.~\cite{Arneodo:1997dp} presents the results for the $Q^2$ dependence of the $^2$H/$^1$H
ratio. In order to reduce a bias due to $Q^2$ dependence, in our analysis 
we use the NMC parametrization of $Q^2$ dependence of their data and select the values
of $Q^2$ close to those of Ref.\cite{Seely:2009gt}.

In order to test the impact of nuclear effects on the extraction of $F_2^n/F_2^p$,
we performed the analysis with the full treatment of nuclear effects
and also with no nuclear effects included, i.e., assuming $R_2=R_3=1$.
Both results are shown in Fig.\ref{fig:npratio}. 
The ratios $R_2$ and $R_3$, used in the full analysis, are shown in Fig.\ref{fig:r2_r3}.
The ratio $R_2$ was calculated as described in Ref.\cite{Kulagin:2004ie}
using the proton and neutron structure functions of Refs.\cite{Alekhin:2007fh,Alekhin:2008ua}
and the Paris deuteron wave function \cite{paris:wf},
while the ${}^3$He structure function and the ratio $R_3$
was calculated by \eq{IA} using the spectral function of Ref.\cite{SS} (see Sec.\ref{sec:spfn}).
We observe from Fig.~\ref{fig:r2_r3} that the EMC effect in $R_3$ is similar to that in $R_2$
with the minimum at $x\sim 0.7$ being somewhat deeper  for $R_3$.
Note that the nuclear effects cancel in the region $x\sim 0.35$,
which is consistent with the measurements of the EMC effect on other nuclei.


\begin{figure}[p]
\begin{center}
\epsfig{file=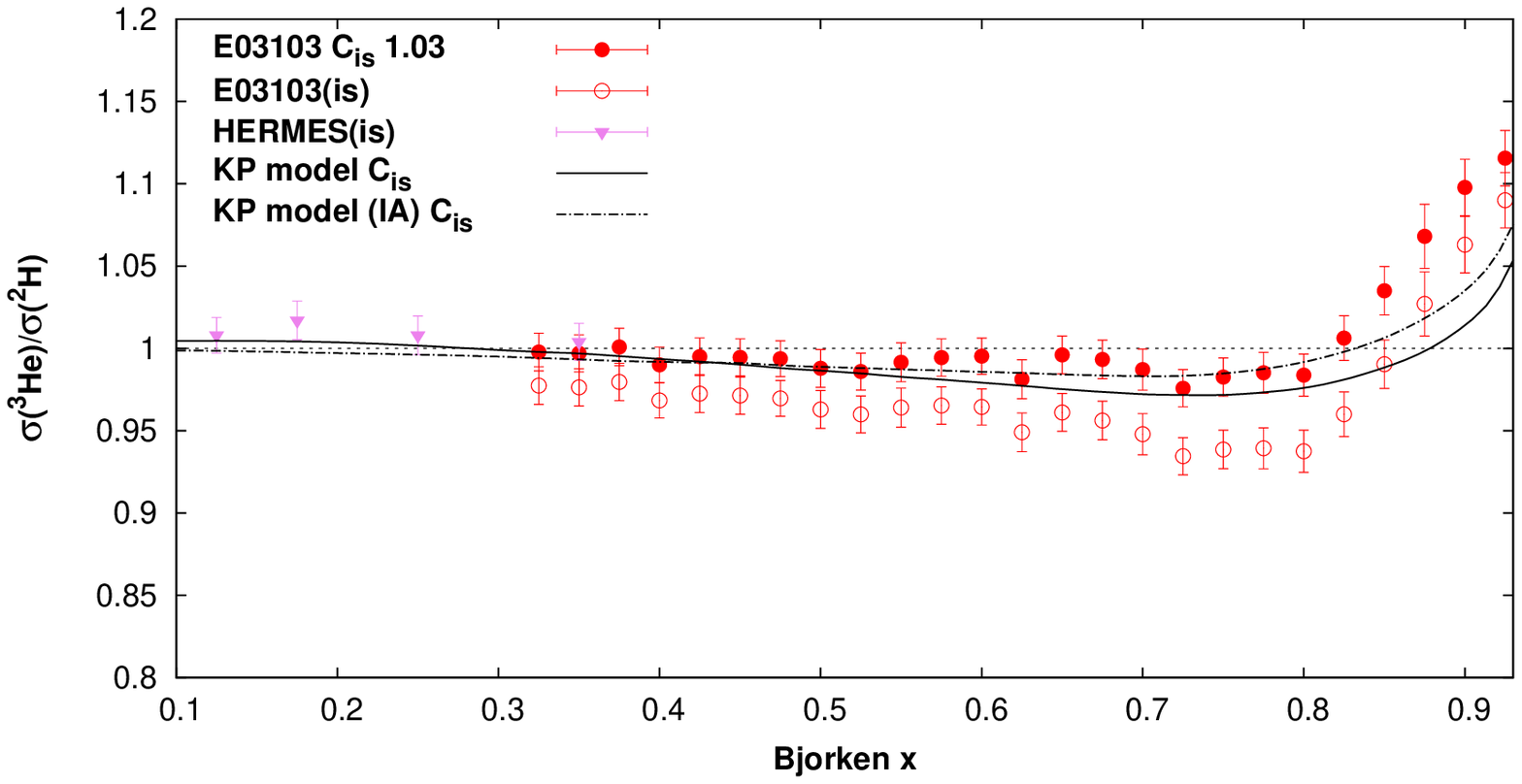,width=\textwidth}
\caption{(Color online)
Ratio of cross sections for electron scattering off ${}^3$He with respect to a 
deuterium target, corrected for the proton excess. 
Data from the HERMES experiment~\cite{Ackerstaff:1999ac} (triangles) 
and the JLab E03-103 experiment~\cite{Seely:2009gt} (open circles) 
are compared with the E03-103 raw data renormalized by a factor $1.03$ and corrected for
the proton excess according to \eq{eq:iso} and using the proton and neutron structure 
functions of Refs.~\cite{Alekhin:2007fh,Alekhin:2008ua} (full circles).  
Our predictions calculated for the same kinematics with the complete model described 
in this article (solid line) and in the impulse approximation (dashed line) 
are also shown for comparison. See text for details.
}
\label{fig:he3is}
\bigskip
\epsfig{file=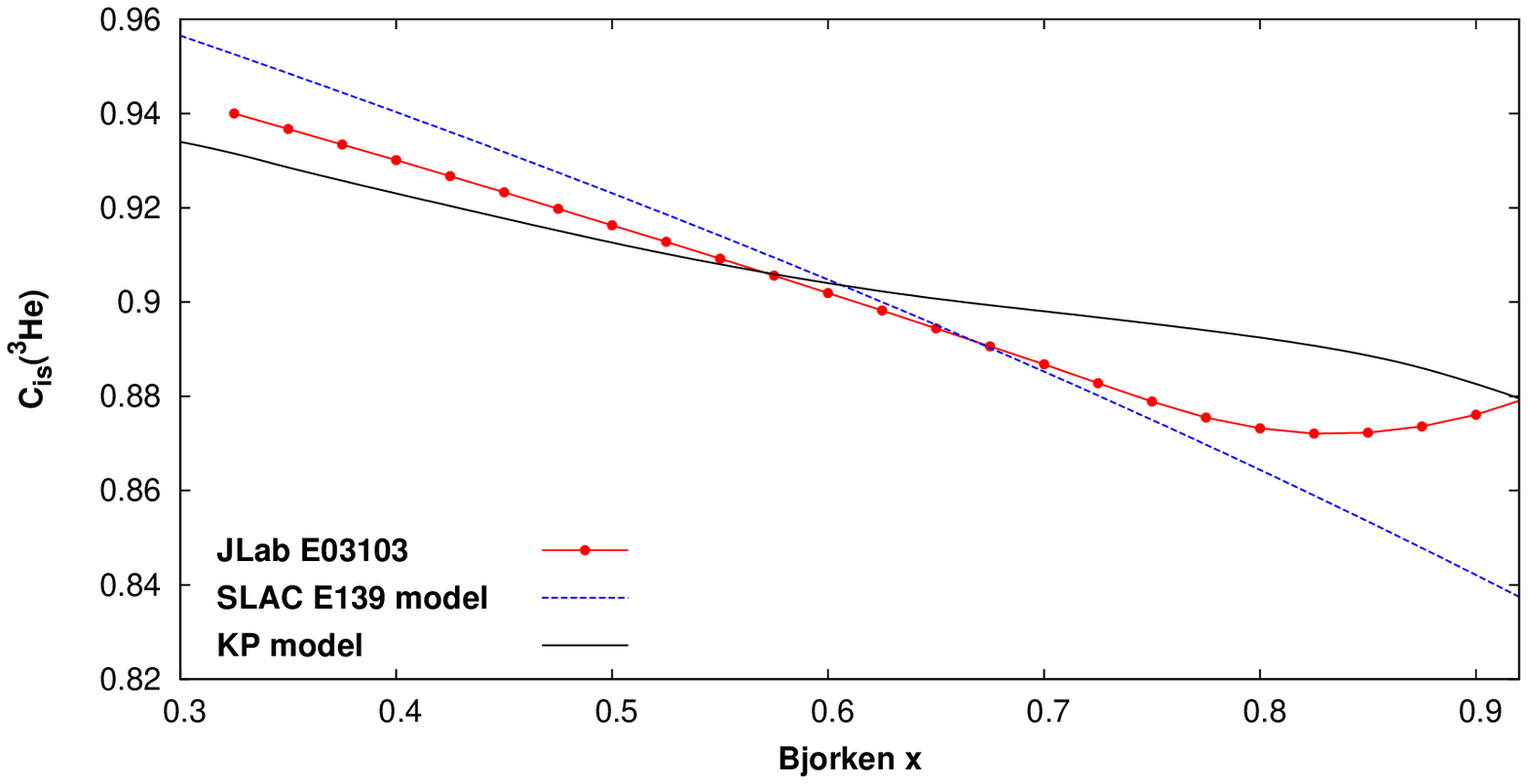,width=\textwidth}
\caption{(Color online)
The isoscalar correction for $^3$He calculated by \eq{eq:iso} using different models
of $F_2^n/F_2^p$.
\label{fig:cis}
}
\end{center}
\end{figure}

A significant mismatch in the values of $F_2^n/F_2^p$  extracted
from different experiments is indicated in Fig.~\ref{fig:npratio}. 
In particular, at $x=0.35$ the central point of the $F_2^n/F_2^p$ ratio from
$^3$He/$^2$H data needs to be rescaled by a factor $0.85^{+0.04}_{-0.03}$ with respect to 
the corresponding value from the $^2$H/$^1$H  data.
In this region the extraction of  $F_2^n/F_2^p$ from nuclear data is practically unaffected
by nuclear effects, as illustrated in Fig.~\ref{fig:npratio}.
Note, however, that for larger $x$ the nuclear corrections are significant.
Curiously enough, the sign of nuclear effect on the ratio $F_2^n/F_2^p$  extracted from  $^3$He/$^2$H 
differs from that extracted from the $^2$H/$^1$H ratio.
This is because the dependence on $R_2$ in \eq{eq:n_p:d} and the dependence on $R_2/R_3$ in \eq{eq:n_p:he3}
essencially differ. 
Note also that in the case of $^3$He/$^2$H the nuclear effects come through the ratio $R_2/R_3$
thus reducing the impact of the Fermi motion at large $x$.

The normalization of  $F_2^n/F_2^p$  is directly related to the normalization of
the $^3$He/$^2$H data and the observed mismatch may signal an inconsistency in the normalization of
nuclear data. 
We apply \eq{eq:n_p:he3} to evaluate necessary renormalization
of $\mathcal R({}^3\mathrm{He}/{}^2\mathrm{H})$ in order to match the $F_2^n/F_2^p$ ratio from $^2$H/$^1$H data and find that
the $^3$He/$^2$H ratio should be increased by a factor $1.03^{+0.006}_{-0.008}$.

In order to further study the nuclear data normalization issue, in Fig.~\ref{fig:he3is} we compare
the $^3$He/$^2$H data from the HERMES~\cite{Ackerstaff:1999ac} and JLab E03-103~\cite{Seely:2009gt}  experiments.
The HERMES data have been corrected for the proton excess using
a correction factor  in \eq{eq:iso} with $F_2^{p,n}$ from the NMC measurements.
The open circles correspond to the ``isoscalar" data as provided in Ref.~\cite{Seely:2009gt}.
The filled circles stand for the E03-103 data corrected for nonisoscalarity in a different way:
We first obtain a raw data by removing the isoscalarity correction of
Ref.~\cite{Seely:2009gt} and then apply the factor $C_{\mathrm is}$ by \eq{eq:iso}
with $F_2^p$ and $F_2^n$ of Ref.~\cite{Alekhin:2007fh,Alekhin:2008ua}
calculated for each $x$ bin.
In addition we also apply a renormalization factor of $1.03$ to each data point.
Also shown are the results of our calculations of $C_{\mathrm is}\mathcal R({}^3\mathrm{He}/{}^2\mathrm{H})$.
It is worth noting that the isoscalarity correction factor $C_{\mathrm is}$ 
differs significantly  from the corresponding correction
in Ref.\cite{Seely:2009gt} and we illustrate its model dependence in Fig.\ref{fig:cis}.
In order to avoid confusion caused by different correction factors, in Fig.~\ref{fig:he3is} 
we compare the calculations with the data corrected in a similar way.
We find from Fig.\ref{fig:he3is} that at the overlap region at $x=0.35$, 
such renormalized E03-103 data are in a good agreement with HERMES data. 

Finally, we note that the nuclear effects in $^3$He are significantly smaller than in
$^4$He (cf. Fig.\ref{fig:e03103} and Fig.\ref{fig:he3is}).
Such a difference is naturally explained in our approach by a dramatic difference
in nuclear binding for these two nuclei, as illustrated in Table~\ref{tab:nuc}.
The off-shell effect is also smaller in $^3$He because its rate is controlled by
the nucleon virtuality $v$ [see \eq{SF:OS}], whose value is driven by nuclear binding.
Using Table~\ref{tab:nuc} we obtain for average virtuality $v=-0.065$\ ($^3$He) and $-0.14$\ ($^4$He).
Thus, a significantly larger value of $|v|$ for $^4$He explains more pronounced off-shell effect
in this nucleus.

\section{Summary} 
\label{sec:sum}

We presented a detailed analysis of nuclear EMC effect for light nuclei.
Focusing on the overlap region $0.1<x<0.7$, we perform a statistical $\chi^2$ analysis of 
the consistency between the data sets from different experiments.   
We found a good agreement between the normalizations of the NMC, SLAC E139 and HERMES experiments 
for nuclei with $A\ge 4$. 
However, the points from the JLab E013-103 experiment appear systematically 
shifted above the SLAC E139 data by an overall normalization factor of $0.98^{+0.005}_{-0.003}$,
common to all discussed nuclei with $A\ge 4$.
This renormalization factor is statistically consistent with the 
normalization uncertainty quoted in Ref.\cite{Seely:2009gt}. After applying the normalization 
offset the JLab E013-103 data are in good agreement with both SLAC E139 and NMC data.

The predictions of Ref.\cite{Kulagin:2004ie} are in a good agreement with renormalized E013-103 data
for all studied nuclei, as discussed in Sec.\ref{sec:comparison}.
We would like to emphasize an important role of off-shell correction,
which is crucial in the description of both the slope of $\mathcal R(x)$ for $0.3<x<0.7$ and position of its minimum
at $x\sim0.75$. This correction is controlled by a product  $\delta f_2(x) v$ averaged with the nuclear
spectral function. The function $\delta f_2(x)$ measures a relative change in the nucleon structure function
due to variation of its invariant mass.
In Ref.\cite{Kulagin:2004ie} this function was phenomenologically derived from a data analysis on nuclear ratios
$\mathcal R$. Further interpretation of observed behavior of $\delta f_2(x)$ in terms of detailed
models would help to better understand mechanisms of modification of the partonic structure of the nucleon
in nuclear environment.


The HERMES data on $^{14}$N are consistent with the NMC data on $^{12}$C at $x>0.1$. 
The predictions by Ref.\cite{Kulagin:2004ie} agree with both experiments. 
We observe that at small $x$ the shadowing effect is more pronounced in the HERMES data 
than in the NMC data. This difference can be attributed to the $Q^2$ dependence, since 
the average $Q^2$ of the HERMES experiment is significantly lower than the corresponding one 
of the NMC experiment. This effect is also confirmed by calculations in a model of 
Ref.\cite{Kulagin:2004ie}.


A significant part of the present analysis is devoted to the study of the EMC effect in $^3$He, 
for which the data from both HERMES and JLab E03-103 experiments are available. In the overlap region at 
$x\sim 0.35$ the normalizations of the two experiments somewhat disagree.
It is important to note that the data on nuclear ratios are usually corrected for the 
proton excess, which depends on the ratio 
$F_2^n/F_2^p$. For this reason, the data appear to be biased by different models 
of the isoscalarity correction. 
In order to verify the consistency of data,
we study the relation between the ratios of nuclear structure functions and $F_2^n/F_2^p$.
We extract $F_2^n/F_2^p$ from both the E03-103 data on the  $^3$He/$^2$H ratio and the NMC data on the  $^2$H/$^1$H ratio.
A significant difference between the results of these two extractions is observed.
In particular, we find that at $x=0.35$ and $Q^2\approx 3\gevsq$ the ratio $F_2^n/F_2^p$ obtained from the JLab E03-103 data
is about 15\% larger than that extracted from the NMC data.
Both extractions of $F_2^n/F_2^p$ become consistent once a normalization factor of 
$1.03^{+0.006}_{-0.008}$
is applied to the $^3$He/$^2$H ratio of the E03-103 experiment.
After such renormalization the E03-103 and HERMES 
data on the ${}^3$He/$^2$H ratio also become consistent,
and our predictions are in  good agreement with both data sets
in the region $x<0.85$.

\section{Acknowledgements} 

We thank J. Arrington, A. Daniel and D. Gaskell for useful communications and information on the E03-103 experiment 
and C. Riedl for useful information on the HERMES data.
We are also grateful to S. Alekhin, W. Melnichouk and S. Scopetta for fruitful discussions 
and communications. 
R.P. thanks USC for supporting this research.



\end{document}